\documentclass{aa}

\usepackage{graphicx}
\usepackage{txfonts}
\usepackage{url}
\usepackage{graphicx}

\usepackage{xcolor}
\usepackage[normalem]{ulem}

\def\kms  {km~s$^{-1}$}

\def\deg  {\ifmmode {^\circ}\else {$^\circ$}\fi}
\def\porm {\ifmmode {\pm}\else {$\pm$}\fi}
\def\chisqpdf {\ifmmode {\chi^2_{\rm pdf}}\else {$\chi^2_{\rm pdf}$}\fi}
\def\chisq    {\ifmmode {\chi^2}\else {$\chi^2$}\fi}

\def\etal {et al.~}

\def\d    {\ifmmode {{\rlap{.}}^\circ}\else {${\rlap{.}}^\circ$}\fi}
\def\s    {\ifmmode {{\rlap{.}}^s}\else {${\rlap{.}}^s$}\fi}
\def\as   {\ifmmode {{\rlap{.}}^{''}}\else {${\rlap{.}}^{''}$}\fi}

\def\pa    {\ifmmode {\psi} \else {$\psi$}\fi}

\def\vlsr  {\ifmmode {v_{\rm LSR}}\else {$v_{\rm LSR}$}\fi}
\def\vlsrr {\ifmmode {v^r_{\rm LSR}}\else {$v^r_{\rm LSR}$}\fi}
\def\vhelio{\ifmmode {v_{Helio}}\else {$v_{Helio}$}\fi}

\def\ura   {\ifmmode {\mu_\alpha}\else {$\mu_\alpha$}\fi}
\def\udec  {\ifmmode {\mu_\delta}\else {$\mu_\delta$}\fi}

\def\ul    {\ifmmode {\mu_l}\else {$\mu_l$}\fi}
\def\ub    {\ifmmode {\mu_b}\else {$\mu_b$}\fi}

\def\uml   {\ifmmode {v_{gr}}\else {$v_{gr}$}\fi}
\def\umb   {\ifmmode {v_b}\else {$v_b$}\fi}
\def\vsrad {\ifmmode {v_{rad}}\else {$v_{rad}$}\fi}

\def\upl   {\ifmmode {v^p_{gr}}\else {$v^p_{gr}$}\fi}
\def\upb   {\ifmmode {v^p_b}\else {$v^p_b$}\fi}
\def\vprad {\ifmmode {v^p_{rad}}\else {$v^p_{rad}$}\fi}

\def\Vo    {\ifmmode {V^{Std}_\odot}\else {$V^{Std}_\odot$}\fi}
\def\Uo    {\ifmmode {U^{Std}_\odot}\else {$U^{Std}_\odot$}\fi}
\def\Wo    {\ifmmode {W^{Std}_\odot}\else {$W^{Std}_\odot$}\fi}
\def\VH    {\ifmmode {V^H_\odot}\else {$V^H_\odot$}\fi}
\def\UH    {\ifmmode {U^H_\odot}\else {$U^H_\odot$}\fi}
\def\WH    {\ifmmode {W^H_\odot}\else {$W^H_\odot$}\fi}
\def\V     {\ifmmode {V_\odot}\else {$V_\odot$}\fi}
\def\U     {\ifmmode {U_\odot}\else {$U_\odot$}\fi}
\def\W     {\ifmmode {W_\odot}\else {$W_\odot$}\fi}

\def\Vs    {\ifmmode {V_s}\else {$V_s$}\fi}
\def\Us    {\ifmmode {U_s}\else {$U_s$}\fi}
\def\Ws    {\ifmmode {W_s}\else {$W_s$}\fi}

\def\Vsbar {\ifmmode {\overline{V_s}}\else {$\overline{V_s}$}\fi}
\def\Usbar {\ifmmode {\overline{U_s}}\else {$\overline{U_s}$}\fi}
\def\Wsbar {\ifmmode {\overline{W_s}}\else {$\overline{W_s}$}\fi}

\def\pars  {\ifmmode{\pi_s}\else{$\pi_s$}\fi}

\def\Ts    {\ifmmode{\Theta_s}\else{$\Theta_s$}\fi}
\def\Tdot  {\ifmmode{d\Theta\over dR}\else{$d\Theta\over dR$}\fi}

\def\Rp    {\ifmmode{R_p}\else{$R_p$}\fi}

\def\To    {\ifmmode{\Theta_0}\else{$\Theta_0$}\fi}
\def\Ro    {\ifmmode{R_0}\else{$R_0$}\fi}
\def\Vlsr {\ifmmode {V_{\rm LSR}} \else {$V_{\rm LSR}$} \fi}

\newcommand{\HII}{\mbox{H\,\textsc{ii}}}%

\definecolor{malachite}{rgb}{0.34, 0.7, 0.22}

\begin{document}

   \title{A comparison of the local spiral structure from Gaia DR2 and
          VLBI maser parallaxes}

   \subtitle{Local spiral structure}

   \author{  Y. Xu\inst{1}, S. B. Bian\inst{1,2}, M. J.
Reid\inst{3}, J. J. Li\inst{1}, B. Zhang\inst{4}, Q. Z. Yan\inst{4},
T. M. Dame\inst{3},
K. M. Menten\inst{5},
Z. H. He\inst{1},
S. L. Liao\inst{4}
\and
Z. H. Tang\inst{4}
          }

   \institute{    Purple Mountain Observatory, Nanjing 210008, China
     \email{xuye@pmo.ac.cn}
     \and
        Shanghai Normal University, Shanghai 200234, China
      \and
      Center for Astrophysics~$\vert$~Harvard \& Smithsonian,
       60 Garden Street, Cambridge, MA 02138, USA
       \and
      Shanghai Observatory, Shanghai 200030, China
      \and
       Max-Planck-Institut f$\ddot{u}$r Radioastronomie,
Auf dem H{\" u}gel 69, 53121 Bonn, Germany}

   \date{Received }
\titlerunning{Local Spiral Structure}
\authorrunning{Xu et al}

  \abstract
   {  The Gaia mission has released the second data set (Gaia DR2),
which contains parallaxes and proper motions for a large number of 
massive, young stars.}
   {  We investigate
the spiral structure in the solar neighborhood  revealed by Gaia DR2 and compare it with  that depicted by VLBI maser parallaxes. }
   {We examined three samples with different constraints on parallax uncertainty and distance errors and stellar spectral types:
(1) all OB stars with parallax errors of less than 10\%;
(2) only O-type stars with 0.1 mas errors imposed and with parallax
distance errors of less than 0.2 kpc; and
(3) only O-type stars with 0.05 mas errors imposed and with parallax
distance errors of less than 0.3 kpc.}
   { In spite of the significant distance uncertainties for stars in DR2 beyond 1.4 kpc,
the spiral structure in the solar neighborhood demonstrated by Gaia
agrees well with that illustrated by VLBI maser results.  The O-type stars available from DR2
extend the spiral arm models determined from VLBI maser parallaxes into the fourth Galactic quadrant, and  
suggest the existence of a new spur between the Local and Sagittarius arms.}
{}

   \maketitle
%


\section{Introduction}

The Milky Way has been thought to be a spiral galaxy since as far back as the 1850s
(Alexander 1852), but it is extremely difficult to observe its spiral structure
due to our edge-on view from its interior and limited by copious dust extinction.
It was not until the 1950s that researchers started to use high-mass stars (OB stars)
and \HII\ regions to trace spiral arm segments in the solar neighborhood \citep{Mor:52,Mor:53}. Such work, augmented by radio wavelength data,  was extended to cover most of the Galaxy by \citet{1976A&A....49...57G} and \citet{2003A&A...397..133R}. 
In the last decade, Very Long Baseline Interferometry (VLBI) at centimeter wavelengths
has provided hundreds of parallaxes to very young, massive stars with accuracies
often as good as $\pm0.01$ mas \citep{Xu:06,Xu:16,Honma:07,Reid:14},
substantially increasing our knowledge of the spiral arms in our Galaxy.
Recent VLBI parallaxes, however,  have mostly been limited to stars visible
from the northern hemisphere, leaving the fourth Galactic quadrant largely
unexplored.

The Gaia satellite \citep{2016A&A...595A...1G}, launched in 2013, will ultimately
achieve parallax accuracy comparable to that of VLBI for approximately $10^9$ stars,
although the current release \citep{DR:2} is limited by systematic error of up to
$<$0.1 mas, depending on celestial positions, magnitudes, and
colors \citep{2018arXiv180409366L}.  Still, even with this limitation,
distances to many OB stars within a couple of kiloparsec from the Sun are now
available.  We note that even when Gaia reaches its target accuracy, direct
measurement of spiral structure will be limited by dust extinction in the
plane of the inner Galaxy to stars typically within a few kiloparsec.

\section{OB star sample}
\label{sec:data}

Owing to their short main sequence lifetimes, high-mass stars are generally located
near their birth places, and consequently can trace Galactic spiral structure.   
However, due to the absence of effective temperatures \citep{DR:2} for bright stars 
(> 10,000 K) in Gaia DR2, we are unable to identify OB stars straightforwardly. 
In order to identify OB stars, we started with a sample of them from the 
``Catalog of Galactic OB Stars'' \citep{2003AJ....125.2531R}, which gives coordinates 
accurate to $\approx1^{''}$,
along with stellar classifications and Heliocentric radial velocities from SIMBAD\footnotemark[1]\footnotetext[1]{ \url{http://simbad.u-strasbg.fr/simbad/sim-fcoo}.}.
We then cross-matched these stars with the Gaia DR2 catalog using a match radius
of $1''$.  After having eliminated those with multiple matches, 5772 O-B2 stars remain
(see Table 1).  Our analysis starts with this sample.


\begin{table*}
\caption{Parallaxes and proper motions of OB stars.}
\label{tab:ob}      
\centering                          
\begin{tabular}{ccllllllll}
        \hline\\
Name & GAIA DR2 ID & R.A.&Dec. & $\pi$ &$\mu_{x}$ & $\mu_{y}$
        &$V_{\rm LSR}$ &Spectral \\
        &  & $(\deg)$& $(\deg )$& (mas) & (mas yr$^{-1}$)&(mas yr$^{-1}$)
        & (km s$^{-1}$)& Type\\
        \hline\\
ALS 13358&528563384392653312&0.1219&67.2168&1.079$\pm$0.051&-1.72$\pm$0.07&-2.39$\pm$0.08&&B1&\\
ALS 13363&423149081478526464&0.2867&58.9799&0.302$\pm$0.030&-3.26$\pm$0.05&-2.62$\pm$0.04&&B0&\\
ALS 13366&423157774492279680&0.3113&59.1428&0.305$\pm$0.028&-3.04$\pm$0.04&-1.34$\pm$0.04&&B2&\\
ALS 13367&429338850909909760&0.3485&60.3555&0.950$\pm$0.054&1.94$\pm$0.09&-3.82$\pm$0.07&-5.2&B2ne&\\
ALS 13375&528594342521399168&0.4453&67.5070&1.016$\pm$0.031&-1.57$\pm$0.04&-1.77$\pm$0.04&&O9.5V&\\
...\\
\hline\\
\end{tabular}
\tablefoot{ Column 1 is the OB star identity in the catalog of \citet{2003AJ....125.2531R}; Column
2 is the unique source identifier in Gaia DR2.  Columns 3 and 4
are barycentric right ascension (R.A.) and declination (Dec.).
Columns 5, 6 and 7 give the parallax and proper motion in the eastward
($\mu_x=\ura \cos{\delta}$) and northward directions ($\mu_y=\udec$).
Column 8 lists local standard of rest (LSR) velocity converted from the
heliocentric value.
Column 9 is the spectral type of the stars.  In Gaia DR2, the
reference epoch is J2015.5. The full table is available in the electronic
attachment.}
\end{table*}

\begin{figure*}
        \centering
        \includegraphics[scale=0.5]{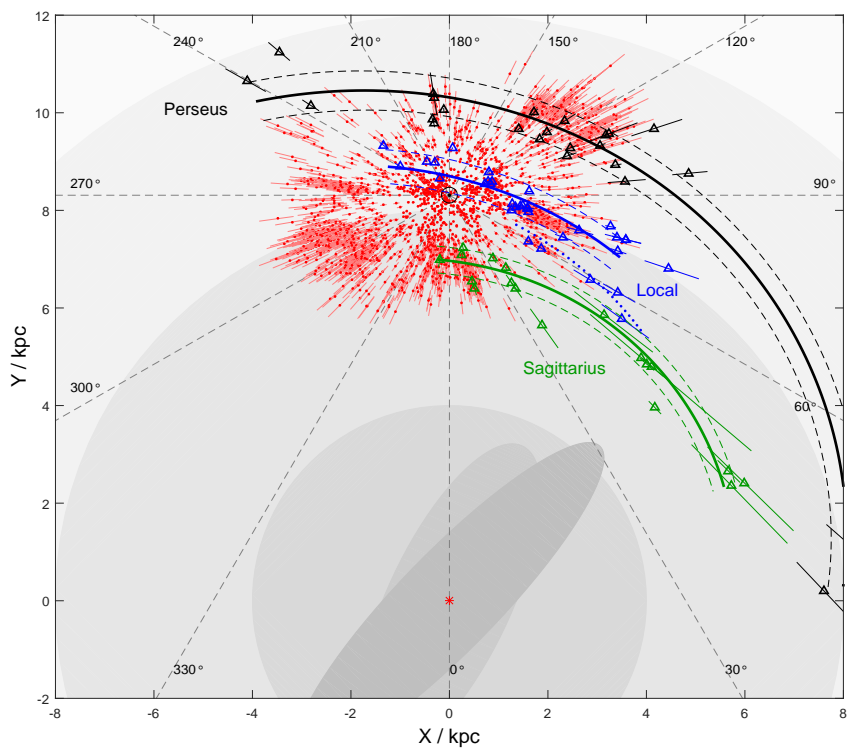}
        \caption{\scriptsize Locations of Gaia  DR2 OB (O - B2) stars (red
circles) and masers (triangles) \citep{Reid:14,Xu:16}.  
The formal parallax uncertainties of the OB stars shown here are better than 10\%,
but do not include a possible $\pm0.1$ mas systematic error. The Galactic center (red
star) is at (0,0) and the Sun (Sun symbol) is at (0, 8.31). The solid and dashed
curved lines denote the center and $\pm1\sigma$ widths of spiral arm models based on
VBLI maser parallaxes.  The Perseus arm (black), the Local arm (blue), and
Sagittarius arm (green) are shown.  Straight dashed lines indicate
Galactic longitude.
 \label{xymap}}
\end{figure*}

\begin{figure*}
        \centering
        \includegraphics[scale=0.5]{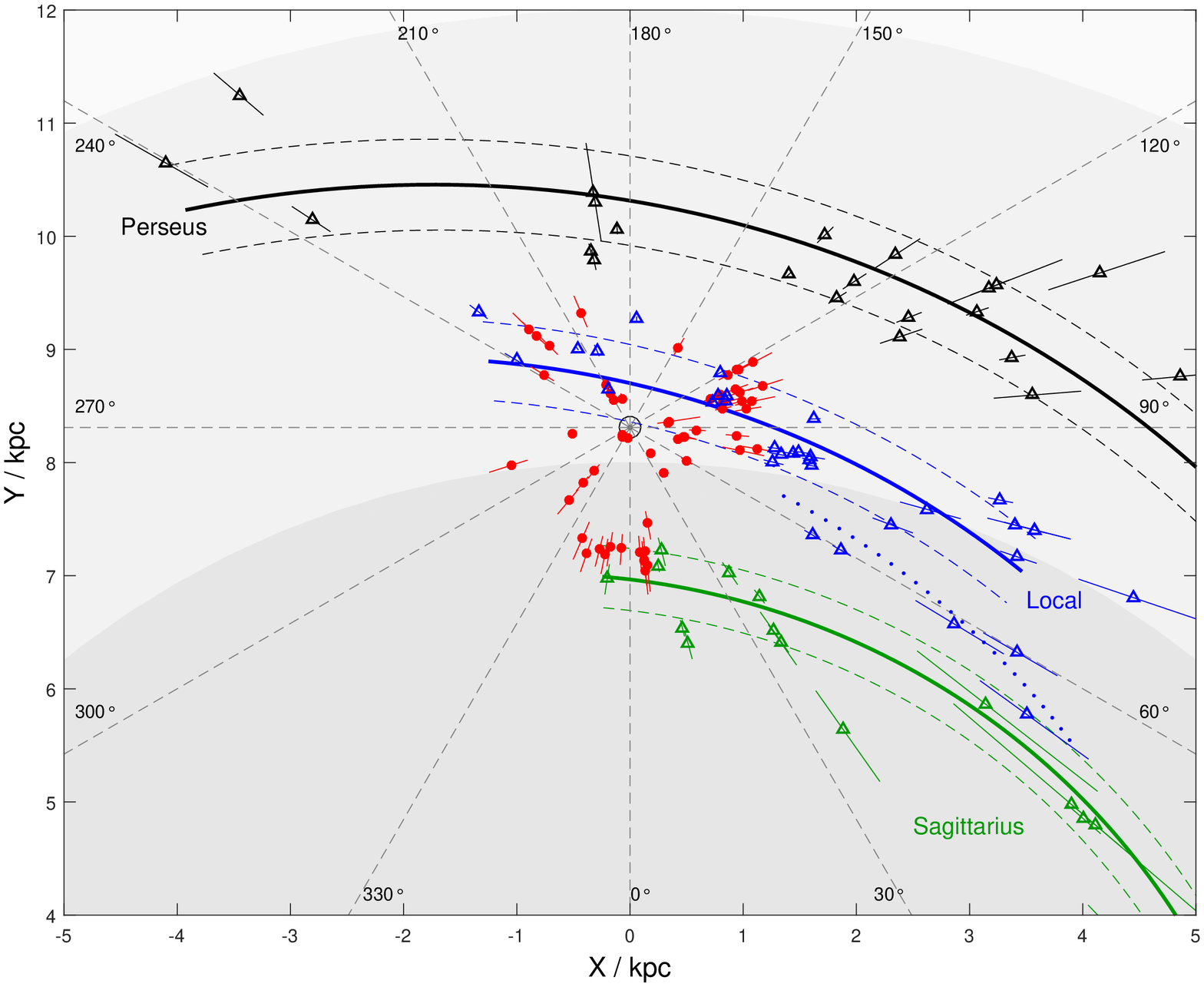}{(a)}
    \centering
        \includegraphics[scale=0.5]{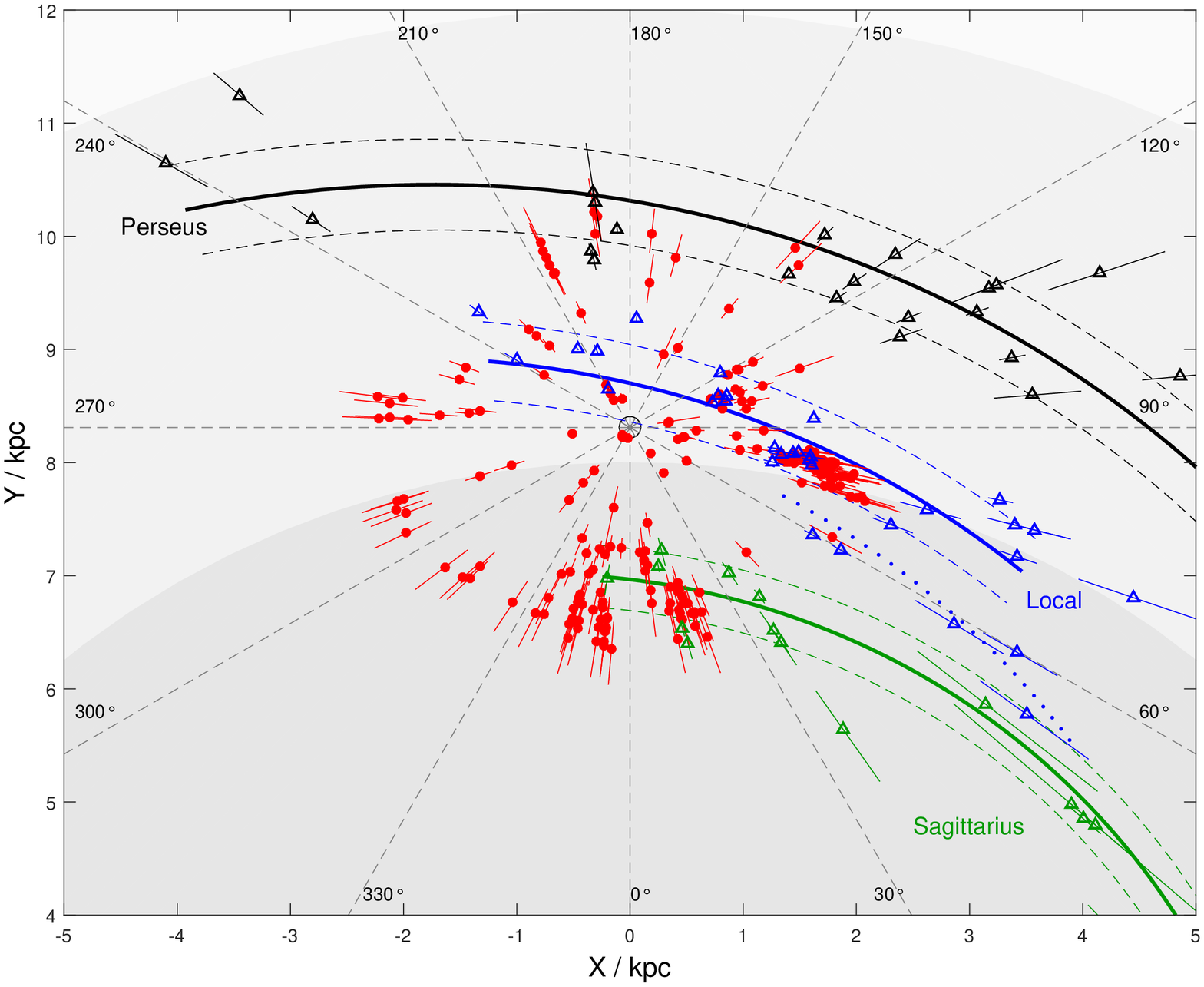}{(b)}
        \caption{\scriptsize Locations of only the identified O-type stars in Gaia DR2 (red circles):
(a) with $\pm0.1$ mas systematic errors added in quadrature to the formal DR2 errors;  only stars with corresponding absolute distance uncertainties $<0.2$ kpc are shown;
(b) with $\pm0.05$ mas systematic errors added in quadrature to the formal DR2 errors; only stars with corresponding absolute distance uncertainties $<0.2$ kpc are shown. 
See Fig. 1 caption for other details, but note that the 10\% error threshold in that
figure is not used here.
}

        \label{fig:arms}
\end{figure*}

Following the recommendation of the Gaia team \citep{2018arXiv180409375A}), we have not corrected the parallaxes of the individual OB stars in our sample by the average systematic bias of 0.029 mas on the parallax zero-point. Such a correction would, in any case, be very small for the nearby OB stars that we examine here. Table~\ref{tab:ob} lists coordinates, parallaxes, proper motions, radial velocities, and classifications of these OB stars. We believe this is the largest sample of O--B2 stars with parallaxes and proper motions to date.

\section{Spiral structure}
\label{sect:Trac}

In this section, we investigate the spiral structure of the Milky Way
within $\approx3$ kpc as revealed by OB stars and compare it with that
determined from maser parallaxes associated with extremely young ($<10^5$ yr) 
high-mass star forming regions, whose parallaxes have been measured accurately with VLBI.
Most of the stars in the sample have parallax uncertainties larger than $\pm20$\%, 
which can lead to distance uncertainty comparable to the spacing between spiral arms.
Therefore, only stars with parallax accuracies better than $\pm10$\% are adopted, yielding
a reduced sample of 2800 sources.  We plot the locations
of these stars in Fig.\ref{xymap}, along with stars with VLBI parallaxes
from \citet{Reid:14} and \citet{Xu:16}, which extend much further into the Galaxy.

As expected, most of the Gaia OB stars in our reduced sample gather around the Sun 
within a radius of about 3 kpc, as distant OB stars have been rejected owing to their 
large distance uncertainties and/or extinctions.
The Gaia data generally confirm the locations of spiral arms from VLBI
in the Perseus, Local, and Sagittarius arms.
Between Galactic longitude 240$^{\circ}$ and 360$^{\circ}$, they can be used
to extend our knowledge of the latter two arms.

As is also apparent in Fig.1, OB stars presumably at nearly the same distance are 
spread out along the line of sight, suggesting that the 10\% threshold for parallax 
uncertainties is insufficient at distances beyond about 1.4 kpc.
In addition, a substantial number of OB stars   appear
to fall between spiral arms.  Since this is rarely the case for the maser sample 
of very young and massive stars, it suggests that allowing stars as late as type 
B2 into the sample is blurring  spiral structure.  These stars can live long enough to 
orbit roughly half the way around the Galaxy and may have left the spiral arms.

In order to obtain an OB star sample  that is truly capable of tracing spiral arms, 
we start over from our initial sample of 5772 stars, abandoning the 10\% uncertainty 
constraint and removing B-type stars.   For the remaining O-type stars,
we incorporate the systematic error 0.1 mas on parallaxes recommended
by \citet{2018arXiv180409366L} and require absolute distance uncertainty
$<0.2$ kpc and peculiar motions $<20$ \kms (the calculations of peculiar motions are given in the appendix).
This leaves a sample of 59 O-type stars.  These stars of O9.7V or earlier 
should have main sequence lifetimes of $<8$ Myr~\citep{2010A&A...524A..98W}, and
with peculiar motions $<20$ \kms\ , they should not have moved more than 0.2 kpc from their
birthplaces.  Because the widths of the spiral arms neighboring
the Sun are $\approx0.3$ kpc \citep{Reid:14} and lines of
sight usually are not perpendicular to spiral arms, these
stars should still be in their birth spiral arms.

In  Fig. \ref{fig:arms}a, we display the distribution of those 59 O-type stars.  Most of them are consistent with being in
the Local and Sagittarius spiral arms, as traced by masers.
Interestingly, about 10 of the stars appear to be between these
two arms.  This observation should be reliable as it comes from a very
conservative treatment of the Gaia DR2 parallaxes.

If we relax the assumed DR2 systematic parallax uncertainty from $\pm0.1$ to $\pm0.05$ mas
and loosen the absolute distance error constraint from $\pm0.2$ kpc to
$\pm0.3$ kpc, the sample increases to 241 stars and is displayed in Fig. 2b.
Significant linear arrangements that point toward the
Sun (``fingers of God'') are now more apparent, suggesting that caution
should be exercised when gleaning information about spiral structure
from this sample.  However, some interesting features, which are not strongly
dependent on the distance uncertainty, can be seen.

The abundance of stars between about 240\deg\ and 270\deg\ longitude at 
distances $>1$ kpc suggest that the Local arm continues into the fourth quadrant 
and bends somewhat inward toward the Galactic center. Similarly, the stars between 
300\deg\ and 320\deg\ longitude at distances $>1$ kpc extend the Sagittarius arm 
into the fourth quadrant with about the same pitch angle as that determined from 
the maser parallaxes.  Finally, a small cluster of stars centered near longitude 
$\approx290$\deg\ at a distance of $\approx2$ kpc appear to extend upward from the 
Sagittarius arm at a fairly high angle, as one might expect for a Galactic spur. 
This should not be surprising as the maser parallaxes have already shown some 
spurs; for example, the Aquila spur \citep{1980ApJ...239L..53C,1986ApJ...305..892D}.


\section{Conclusions}

VLBI parallaxes have traced the Milky Way's spiral structure within
about 10 kpc of the Sun in Galactic quadrants 1, 2, and 3. The Gaia
DR2 data supplement the maser VLBI measurements and extend
the map of the Local and Sagittarius arms into quadrant 4.
They also suggest an additional spur-like feature between those arms.

\begin{acknowledgements}This work was supported by the National Natural Science Foundation of
China (Grant Numbers: 11673066, U1431227, 11673051, 11703065, 11573054, 11503042), the Natural
Science Foundation of Shanghai under grant 15ZR1446900, and 
the 100 Talents Project of the Chinese Academy of Sciences. This work has made use of data from the European Space Agency (ESA)
mission
{\it Gaia} (\url{https://www.cosmos.esa.int/gaia}), processed by
the {\it Gaia}
Data Processing and Analysis Consortium (DPAC,
\url{https://www.cosmos.esa.int/web/gaia/dpac/consortium}). Funding
for the DPAC
has been provided by national institutions, in particular the institutions
participating in the {\it Gaia} Multilateral Agreement.
\end{acknowledgements}


%
%

\section{APPENDIX}\label{online}

\subsection{The calculation of the peculiar motions of O stars}

With distances, proper potions, and radial velocities, one has full
three-dimensional (3D) velocity information in a heliocentric reference
frame.  Adding the Sun's full motion with respect to a reference
frame at rest at the Galactic center transfers the O-star motions
to that frame.  Then removing a model of pure circular motion gives
Galactic peculiar motions, expressed by $(U_{s},V_{s},W_{s})$,
which are velocity components toward the Galactic center (GC), in
the direction of Galactic rotation, and toward the north Galactic
Pole at the location of each star.

We estimate peculiar motions of the O stars following \citet{Reid:09},
using updated Galactic parameters of 241~\kms\ for the Galactic rotation
speed, $\Theta_{0}$, at a distance of 8.31 kpc to the GC, $R_{0}$,
and solar motion parameters of $U_{\odot} = 10.5$ \kms,
$V_{\odot} = 14.4$ \kms, and $W_{\odot} = 8.9$ \kms\ from \citet{Reid:14}.
We used the ``Universal'' rotation curve model, which
produces a relatively flat rotation curve between 5 and 15 kpc.

\end{document}